\documentstyle[prl,aps,twocolumn,epsf]{revtex}
\def\break#1{\pagebreak \vspace*{#1}}
\begin{document}


\title{Gravitational bouncing of a quantum ball: Room for Airy's function Bi}

\author{
Haret C. Rosu
}

\address{ 
Instituto de F\'{\i}sica de la Universidad de Guanajuato, Apdo Postal
E-143, Le\'on, Guanajuato, M\'exico}  

\maketitle
\widetext

\begin{abstract}
I apply (i) a classical version of the Ermakov-Lewis procedure and (ii) 
the strictly isospectral
supersymmetric approach to the Schroedinger free fall of the 
bouncing ball type. In both cases, 
the Airy function Bi, which in general is eliminated as being unphysical, 
plays a well-defined role. Relevant plots are displayed.
\end{abstract}
\vskip 0.1in

PACS number(s):  03.65.-w, 11.30 Pb 
\vskip 0.1in


\narrowtext

\section*{I. Introduction} 
Bouncing of a cold atomic cloud has been first observed in the 
laboratory in 1993 \cite{fr}.
It has been shown that cold atoms dropped onto an ``atomic mirror" can be used 
for holographic manipulation of atomic beams \cite{jap}. More recently, 
bouncing Bose-Einstein condensates
have been examined in the laboratory \cite{bbec}.
Here, we consider the toy model of a Schroedinger quantum particle bouncing on a 
perfectly reflecting surface in a linear gravitational field, which is known 
as the quantum bouncing ball [QBB] problem \cite{gb99}. 
In the QBB case, one should solve the Schroedinger equation with the potential 
\begin{eqnarray}
V_{\rm QBB}(z) &=mgz&, \qquad  \mbox{if $z > 0$} \nonumber \\
V_{\rm QBB}(z) &=\infty & , \qquad \mbox{if $z\leq 0$~.}
      \end{eqnarray}
By the scalings ${\rm s}=z/l_g$ and ${\rm S} =E/mgl_g$, where
$l_g=\left(\frac{h^2}{2m^2g}\right)^{1/3}$ is 
the ``gravitational length" unit, the stationary QBB Schroedinger equation 
becomes dimensionless
\begin{equation}
{\rm \frac{d^2\psi}{ds^2}-(s-S)\psi} =0~.
\end{equation}
The general solution is a superposition of Airy functions Ai(s) and Bi(s), 
but Airy's Bi is discarded for going to infinity at large ${\rm s}$. Moreover,
the perfectly reflecting boundary requires the wave function be zero
at the origin and therefore the physical eigenmodes are written as
${\rm \psi _n(s)=N_n{\rm Ai}(s-S_n)}$, where ${\rm N_n}$ is the normalization 
constant and $\rm S_n$ are the zeros of the Ai function \cite{gb99}. 
In other words, a shift of the Airy's argument is performed placing
the Airy zeros at the origin. To the best of the author's knowledge all the 
previous works in this field made use of only Airy function Ai of 
shifted argument. The main purpose here 
is to show that there are two techniques in which the Bi function could
still be employed without leading to unphysical results. 
One of them is the Ermakov-Lewis (EL) procedure, which is presented 
in section II and the other 
one is the strictly isospectral supersymmetric (SUSY) approach enclosed 
in section III. A small conclusion section ends up the work.  
\break{0.70in}

\section*{II. Classical Ermakov-Lewis Approach for QBB}
I will use the version of the EL approach \cite{EL}
that I introduced in 
previous works in collaboration \cite{colab}. Eq.~(2) can be mapped in a 
known way 
to the canonical equations for a classical point particle of 
unit mass, generalized coordinate ${\rm q=\psi}$,
momentum ${\rm p=\dot{\psi}}$, (i.e., velocity $v=\dot{\psi}$), 
where the dot means 
total derivative with respects to ${\rm s}$, i.e., we
identify the coordinate $s$ with the classical Hamiltonian time.
Thus, one is led to
\begin{eqnarray}
\rm \dot{q}\equiv\frac{dq}{ds}&=&\rm p~\\   
\rm \dot{p}\equiv
\frac{dp}{ds}&=&\rm (s-S)q~.
\end{eqnarray}
These equations describe the canonical
motion for a classical point particle as derived from the
time-dependent Hamiltonian of the inverted oscillator type  
\begin{equation} \label{4}
{\rm H_{\rm cl}(s)=
\frac{p^2}{2}-(s-S)\frac{q^2}{2}}~.
\end{equation}
For this classical Hamiltonian the triplet of phase-space
functions $T_1={\rm \frac{p^2}{2}}$, $ T_2={\rm pq}$,
and $ T_3={\rm \frac{q^2}{2}}$ forms a dynamical Lie algebra, i.e.,
${\rm H_{\rm cl}}= {\rm \sum _{n=1}^{3}h_{n}(s)}T_{\rm n}({\rm p,q})$, 
which is closed with respect to the Poisson bracket, namely
$ \{T_1,T_2\}=-2T_1$, $\{T_2,T_3\}=-2T_3$, $\{T_1,T_3\}=-T_2$. 
Using this algebra ${\rm H_{\rm cl}}$ reads
\begin{equation} \label{H}
{\rm H_{\rm cl}}=T_1-{\rm (s-S)} T_3~.
\end{equation}
The Lewis invariant ${\cal I}$ belongs to the dynamical algebra, i.e.,
one can write 
${\rm {\cal I}(s)=\sum _{r}\epsilon _{r}(s)}T_{\rm r}$, 
and by means of
${\rm \frac{\partial {\cal I}}{\partial s}=-\{{\cal I},H\}}$ one is led to the
following equations for the functions ${\rm \epsilon _{r}(s)}$
\begin{equation} \label{7}
{\rm \dot{\epsilon} _{r}+\sum _{n}\Bigg[\sum _{m}C_{nm}^{r}h_{m}
(s)\Bigg]\epsilon _{n}=0}~,
\end{equation}
where ${\rm C_{nm}^{r}}$ are the structure constants of the Lie algebra that
have been already given above. Thus, we get
\begin{eqnarray} \nonumber
\rm \dot{\epsilon} _1&=&-2\epsilon _2 \\
\dot{\epsilon} _2&=&\rm -(s-S)\epsilon _1-\epsilon _3\\
\dot{\epsilon} _3&=&\rm -2(s-S)\epsilon _2~. \nonumber
\end{eqnarray}
The solution of this system can be readily obtained by setting
${\rm \epsilon _1=\rho ^2}$ giving ${\rm \epsilon _2=-\rho
\dot{\rho}}$ and
${\rm \epsilon _3=\dot{\rho} ^2 +
\frac{1}{\rho ^2}}$, where $\rho$ is the solution of the Milne-Pinney (MP)
equation \cite{P},
${\rm \ddot{\rho}-(s-S) \rho=\frac{1}{\rho ^3}}$.
Since Pinney's note in 1950 it is widely known how to
write $\rho$ as a
function of the two particular solutions of the corresponding parametric
oscillator problem.  
We have followed the method of Eliezer and Gray \cite{EG}
in order to write ${\rm \rho(s)}$ as a combination 
of Airy functions 
that satisfy the initial conditions as given by those authors.
Thus, we used in all our calculations the following formula
\begin{equation} \label{rho}
{\rm \rho  _{1}(s)=N_1\left[ (Ai(s-S_1)+Bi(s-S_1))^2+ 
Bi^2 (s-S_1)\right] ^{1/2}}~,
\end{equation}
i.e., we used the two Airy functions corresponding to the ground state.
In Eq.~(9), $\rm S_1=(9\pi/8)^{2/3}$ and 
$\rm N_1=(8\pi ^2/9)^{1/6}$ \cite{gb99}.
In terms of the MP solution ${\rm \rho (s)}$ the Lewis invariant reads 
$$
{\rm {\cal I} _{n}(s)=\frac{(\rho _{n} p-\dot{\rho_n}q)^2}{2}
+\frac{q^2}{2\rho _{n} ^2}}
$$
\begin{equation} \label{10}
=
{\rm \frac{1}{2}\left(
\rho _{n} \dot{\psi} _{n}-\dot{\rho _n} \psi _{n}\right)^2
+\frac{1}{2}\left(\frac{\psi _{n}}{\rho _{n}}
\right)^{2}}~.
\end{equation}
For example, one can check by direct calculation 
that ${\rm {\cal I} _{1}(s)}=\frac{1}{2}$ since 
according to the Eliezer-Gray interpretation the EL invariant should be 
$\frac{1}{2}h^2$ where $h$ is the coefficient of the inverse cubic nonlinearity
in the aforewritten MP equation where $h=1$.

In the EL approach the angular quantities are given by the following formulas
\cite{mor}
\begin{equation}
\Delta \theta ^{{\rm d}}=
\int _{0}^{T}\Big[\frac{1}{\rho ^2}-
\frac{1}{2}\frac{d}{ds ^{'}}
(\dot{\rho}\rho)+\dot{\rho}^2\Big]d s ^{'}
\end{equation}
and
\begin{equation}
\Delta \theta ^{{\rm g}}=\frac{1}{2}\int _{0}^{T}
\Big[\frac{d}{d s ^{'}}
(\dot{\rho}\rho)-2\dot{\rho}^2\Big]ds ^{'},
\end{equation} 
for the dynamical and geometrical angles, respectively.
Thus, the total angle will be
\begin{equation}
\Delta \theta ^{{\rm t}}=\Delta \theta ^{{\rm d}}+ \Delta \theta ^{{\rm g}}
=\int _{0}^{T}\frac{1}{\rho ^2}ds ^{'}~.
\end{equation}
Plots of all these angles calculated using $\rho _1$ 
are displayed in Figs~1,2,3, respectively. 

\section*{III. Strictly isospectral SUSY bouncing ball} 
Factorizations of one-dimensional Schroedinger operators have been  first
discussed in the SUSY context
by Witten in 1981 \cite{w81}, and are well known in the mathematical
literature in the
broader sense of Darboux covariance of Schroedinger equations
\cite{d82}.

In 1984, Mielnik \cite{mn} introduced a different factorization of the
quantum harmonic oscillator based on the general Riccati solution 
here denoted by $\rm w_g$. As a result,
Mielnik obtained a one-parameter
family of potentials with {\em exactly} the same spectrum as that of the
harmonic oscillator. 
Mielnik's method offers  an interesting possibility to construct families of 
potentials
{\em strictly} isospectral with respect to the initial (bosonic) one
by simply taking into account the most general superpotential (i.e., the general
Riccati solution). Thus, in the QBB case one requires
$\rm V_+(s)=  w_{g}^2 + \frac{d w_{g}}{ds}$, where ${\rm V_+}$ is
the fermionic partner potential of $V_{\rm QBB}$. It is easy to see that
one particular solution to this equation is ${\rm w_p= w(s)}$, where 
$\rm w(s)=-\psi _{1} ^{\prime}/\psi _1$ is
the common Witten superpotential. One is led to consider the following
Riccati equation ${\rm  w_{g}^2 + \frac{d w_{g}}{ds}=w^2_p +\frac{d w_p}{ds}}$,
whose general solution can be written down as 
${\rm w_{g}(s)= w_p(s) + \frac{1}{v(s)}}$, where ${\rm v(s)}$ is an unknown
function. Using this ansatz, one obtains for the function ${\rm v(s)}$ the
following Bernoulli equation
\begin{equation}
{\rm \frac{dv(s)}{ds} - 2 \, v(s)\, w_p(s) = 1},
\end{equation}
that has the solution
\begin{equation}
{\rm v(s)= \frac{I_0(s)+ \lambda}{\psi_{1}^{2}(s)}}~.
\end{equation}
The integral ${\rm I_0(s)= \int _{0}^{s} \, \psi_1^2(y)\, dy}$ is a 
step-like function as one can see in Fig.~4. On the other hand, 
$\lambda >0$ is an
integration constant thereby considered as a free parameter, 
which is a measure 
of the contribution of the second linearly independent solution, i.e., the
Airy Bi in the QBB case, as we argued elsewhere \cite{boya}.
Thus, ${\rm w_{g}(s)}$ can be written as follows
$$
{\rm w_{g}(s;\lambda)=  w_p(s) + \rm \frac{d}{ds}} \Big[ {\rm ln
(I_0(s) + \lambda) \Big]}
$$
\begin{equation}
={\rm - \frac{d}{ds} \Big[ ln \left(\frac{\psi_1(s)}{I_0(s) +
\lambda}\right)\Big]}.
\end{equation}
Finally, one easily gets the parametric family of 
potentials 
$$
{\rm  V(s;\lambda)} = {\rm w_{g}^2(s;\lambda) -
\frac{d w_{g}(s;\lambda)}{ds}}
$$
$$
= {\rm V_{QBB}(s) - 2 \frac{d^2}{ds^2} \Big[ ln(I_0(s) + \lambda)}
\Big]
$$
\begin{equation}
= {\rm V_{QBB}(s) - \frac{4 \psi _1(s) \psi _1^\prime (s)}{I_0(s)
+ \lambda} 
+ \frac{2 \psi _1^4(s)}{(I_0(s) + \lambda)^2}}~.
\end{equation}
All ${\rm V(s;\lambda)}$ have the same SUSY partner potential
${\rm V_+(s)}$ obtained by deleting the ground state.
They may be considered as a sort
of intermediates between the bosonic potential $ {\rm V_{QBB}(s)}$ and
the fermionic counterpart ${\rm V_+(s)}$.  
A plot of ${\rm V(s;\lambda)}$ is given in Fig.~5.
From Eq. ($16$) one can infer the ground state wave functions
for the potentials ${\rm V(s;\lambda)}$ as follows
\begin{equation}
{\rm  \varphi _1(s;\lambda)= N(\lambda)
\frac{\psi _1(s)}{I_0(s) + \lambda}}~,
\end{equation}
where $\rm N(\lambda)$ is a normalization factor that can be shown to be
of the form $\rm N(\lambda)= \sqrt{\lambda(\lambda +1)}$. The normalized 
functions ${\rm \psi _1}$
and ${\rm \varphi _1}$ are plotted in Fig.~6.

\section*{IV. Conclusion} 

Airy's function ${\rm Bi}$ can find a place in the physics 
of the quantum bouncing ball through two theoretical procedures
connecting the Schroedinger equation with the nonlinear
Milne-Pinney equation and Riccati equation, respectively.
This may help in gaining further insight in the problem of nonrelativistic 
quantum free fall. The Lewis angles and phases that depend on the function 
${\rm Bi}$ through the Milne-Pinney function are important quantities 
provided by the Ermakov-Lewis approach that is here 
applied to a Schroedinger free fall problem for the first time. 
These quantities are similar to
Berry phases and Hannay angles and in principle can be measured in 
quantum bouncing ball experiments. On the other hand, 
in the strictly isospectral supersymmetric
method, the contribution of the ${\rm Bi}$ function enters through the 
parameter $\lambda$ \cite{boya}. However, although the results are 
physically sound, it is still not clear what is the 
corresponding experimental configuration. In other words, it is not clear 
how a strictly isospectral partner potential, such as the one displayed 
in Fig. 5 can 
be produced experimentally. For example, one may think of some particular 
microscopic surface effects of the atomic mirror that might be able to
distort the interaction potential in the way the SUSY scheme 
predicts.




\newpage

\vskip 1ex
\centerline{
\epsfxsize=220pt
\epsfbox{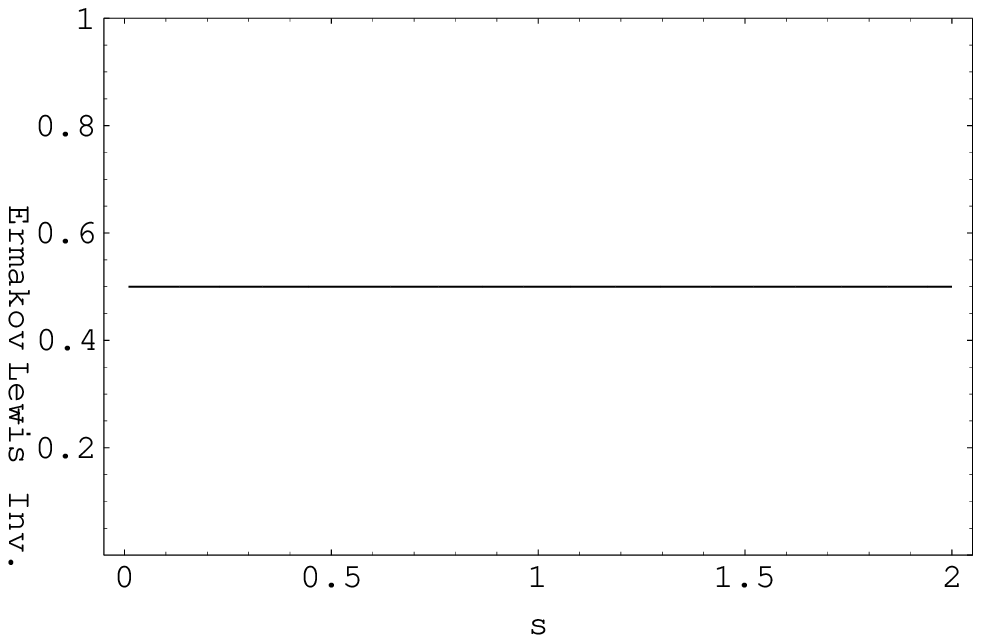}}
\vskip 3ex
\begin{center}
{\small{FIG. 0}.
Ermakov-Lewis invariant ${\cal I} _{1}({\rm s})$} cf. Eq.~(10) [not in the 
accepted version.
\end{center}


\centerline{
\epsfxsize=200pt
\epsfbox{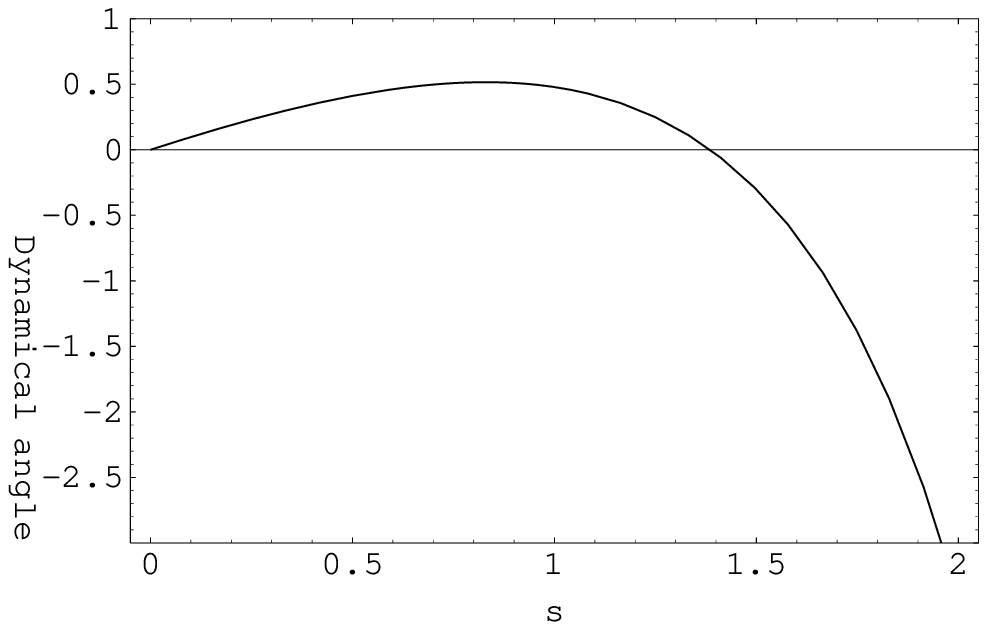}}   
\vskip 3ex
\begin{center}
{\small{FIG. 1}.
Lewis' dynamical angle cf. Eq.~(11). 
}
\end{center}

\vskip 1ex
\centerline{
\epsfxsize=200pt
\epsfbox{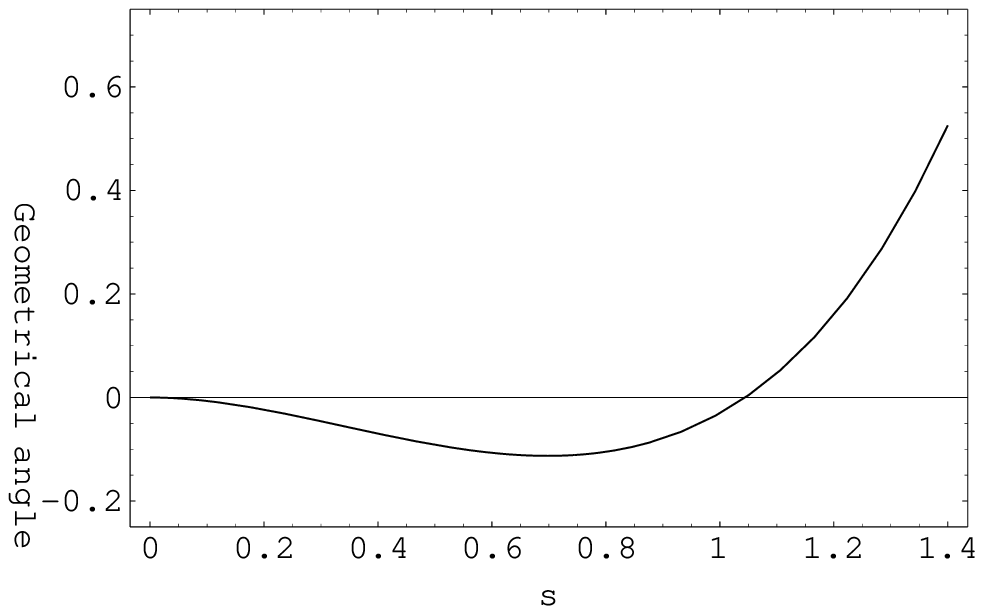}} 
\vskip 3ex
\begin{center}
{\small{FIG. 2}.
Lewis' geometric angle cf. Eq.~(12).}
\end{center}

\newpage

\vskip 1ex
\centerline{
\epsfxsize=210pt
\epsfbox{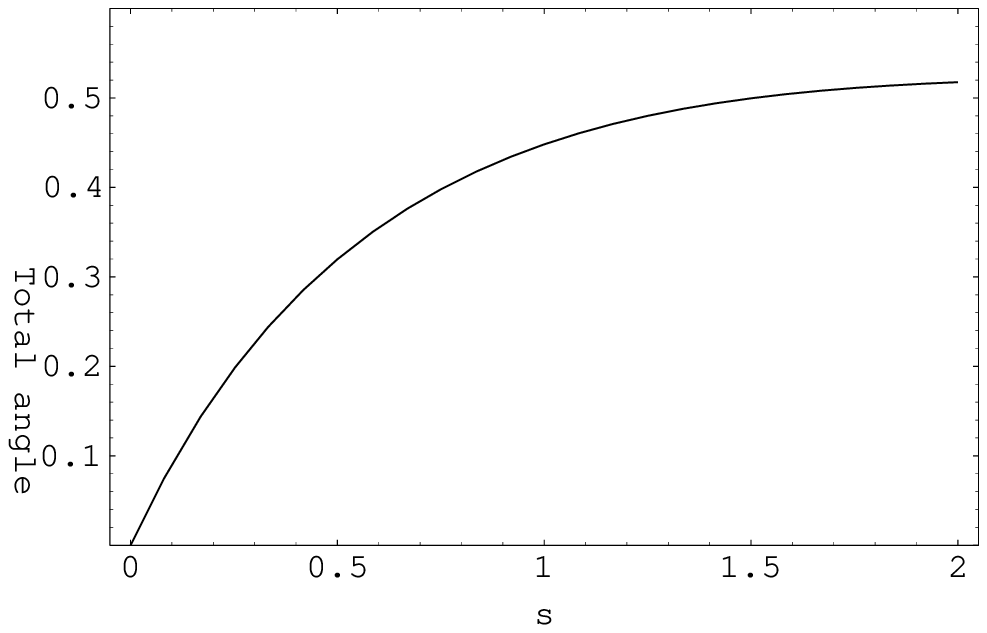}}  
\vskip 3ex
\begin{center}
{\small{FIG. 3}.
Lewis' total angle cf. Eq.~(13).}
\end{center}

\vskip 1ex
\centerline{
\epsfxsize=220pt
\epsfbox{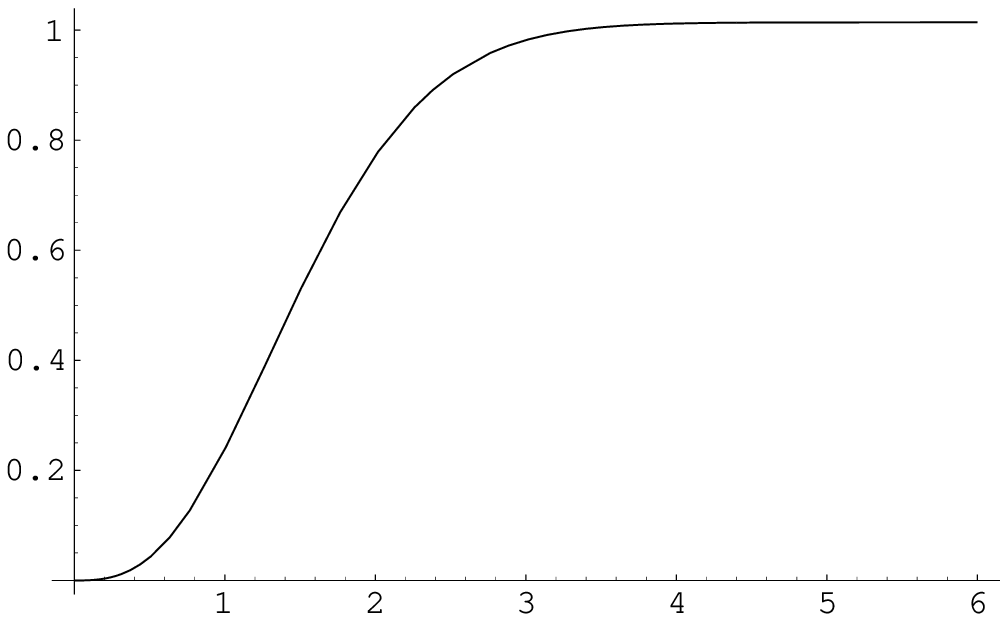}}
\vskip 3ex
\begin{center}
{\small{FIG. 4}.
 The integral ${\rm I_0(s)}$ of the strictly isospectral SUSY QBB. }
\end{center}

\vskip 1ex
\centerline{
\epsfxsize=220pt
\epsfbox{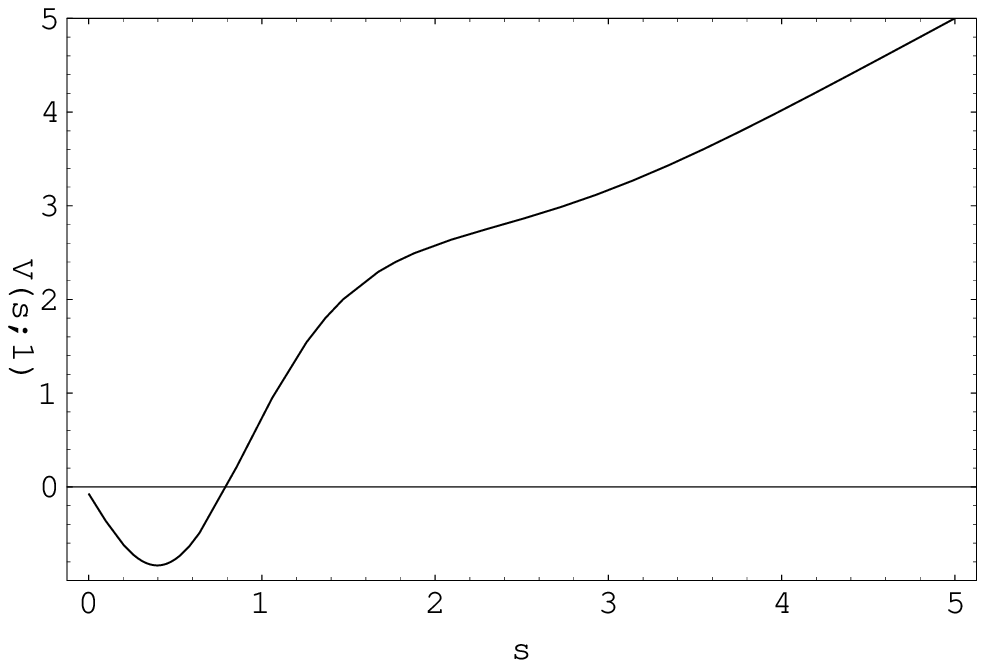}}  
\vskip 3ex
\begin{center}
{\small{FIG. 5}.
 The strictly isospectral QBB gravitational potential for $\lambda =1$.}
\end{center}

\vskip 1ex
\centerline{
\epsfxsize=210pt
\epsfbox{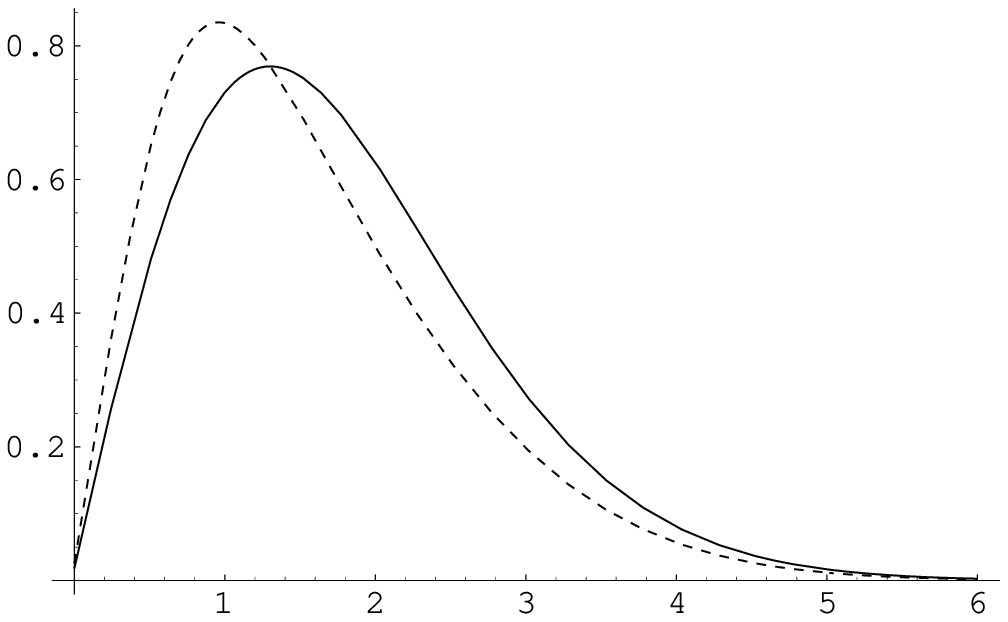}}
\vskip 3ex
\begin{center}
{\small{FIG. 6}.
The normalized wave functions ${\rm \psi _1(s)}$ (full line) and 
${\rm \varphi _1(s; 1)}$ (dashed line).}

\end{center}


\end{document}